# An Operational Data-Driven Malfunction Detection Framework for Enhanced Power Distribution System Monitoring – The DeMaDs Approach

David Fellner
AIT Austrian Institute
of Technology – Austria
David.Fellner@ait.ac.at

Thomas I. Strasser
AIT Austrian Institute of Technology
& TU Wien – Austria
Thomas.I.Strasser@ieee.org

Wolfgang Kastner
Technische Universität Wien
(TU Wien) – Austria
Wolfgang.Kastner@tuwien.ac.at

Behnam Feizifar
University of Strathclyde – United Kingdom
Behnam.Feizifar@strath.ac.uk

Ibrahim F. Abdulhadi
University of Strathclyde – United Kingdom
Ibrahim.F.Abdulhadi@strath.ac.uk

## ABSTRACT

*The changes in the electric energy system toward a sustainable future are inevitable and already on the way today. This often entails a change of paradigm for the electric energy grid, for example, the switch from central to decentralized power generation which also has to provide grid-supporting functionalities. However, due to the scarcity of distributed sensors, new solutions for grid operators for monitoring these functionalities are needed. The framework presented in this work allows to apply and assess data-driven detection methods in order to implement such monitoring capabilities. Furthermore, an approach to a multi-stage detection of misconfigurations is introduced. Details on implementations of the single stages as well as their requirements are also presented. Furthermore, testing and validation results are discussed. Due to its feature of being seamlessly integrable into system operators' current metering infrastructure, clear benefits of the proposed solution are pointed out.*

## INTRODUCTION

As the energy infrastructure as a whole takes a necessary shift to operate sustainably, the electrical energy system is strongly affected as well. Especially the introduction of decentralized renewable energy generation, but also other novel grid-connected devices lead to new challenges in grid operation. The problems raised originate from introducing new grid participants such as Photovoltaic (PV) inverters, Battery Energy Storage Systems (BESS), or Electric Vehicle Supply Equipment (EVSE). They lead to unseen consumption patterns but also to bidirectional power flows which the distribution grid was not designed for. The said devices are increasingly important for maintaining grid stability as they provide grid-supporting functionalities [1]. These functionalities help, for example, to keep the voltage in defined boundaries by dispatching reactive power in the case of PV inverters or prevent overloading by controlling the charging currents of EVSEs [2]. As these control features are implemented locally, a deviation from the specified configuration might go unnoticed. Settings might reset to default values, due to errors. The configuration might also change to other settings as a result of improper handling of devices. In such a case, the respective grid-supporting functionalities are no longer provided in the desired manner. This can pose a problem for the safe and reliable operation of the grid. Especially cases of more than one device being affected are to be avoided so as not to let their impacts accumulate. Distribution System Operators (DSOs), therefore, need a novel tool to monitor their low-voltage power distribution grids to detect misconfigured devices. This tool ought to work under the limitations set by the scarce proliferation of sensors in the low-voltage grid, as well as the common lack of trained staff. Furthermore, data protection issues have to be taken into account. They limit the use of smart meter readings, especially by a central control entity. Moreover, easy installation and roll-out are important factors to protect the scalability of such as solution. Due to these constraints, a data-driven solution is generally promising. However, the state-of-the-art does not provide much insight into what such a solution could look like. Existing works using Machine Learning (ML) are either targeting the detection of faults or disturbances on a system level [3] or the detection of attacks on the electric power grid [4]. Both approaches are not suited for detecting misconfigurations on a device level or in a grid segment.

Therefore, the subject addressed is the increased need of DSOs for monitoring low-voltage distribution grids. A framework providing a suitable solution is presented in this work that uses generally available grid operational data, such as voltages and currents, to recognize a misconfiguration present in grid-supporting control functions of certain devices, such as inverter-based Distributed Energy Resources (DER).

The paper is structured as follows: Section 'Introduction' provides the motivation and background for the presented work followed by Section 'Framework Architecture' which lays out the structure of the development framework and the solution developed. Section 'Validation Results' presents details on the implementation of the solution and the performance achieved with regard to detection. This includes the performance of data collected in a lab setting. Finally, Section 'Conclusions' summarizes the work and points out possibilities for future extensions of the proposed framework.

## FRAMEWORK ARCHITECTURE

The proposed solution is sketched in Figure 1. Measurements at the secondary substation of a grid as well as at the grid participants' connection points are used. The data at the grid connection points are either collected through smart meters or gained through data mining of the substation level data. The data from both sources are fed into the DeMaDs (Data Driven Detection of Malfunctioning Devices in Power Distribution Systems) framework. If the data stems from local meter measurements, the usage of this data for detection purposes is solely kept on a decentral level. The DeMaDs framework then uses data-driven methods such as ML or Deep Learning (DL) [5] to build a classifier. This classifier is trained to distinguish between regular operational states, in which the grid participants show the intended behavior, and erroneous states. In the latter case, a misconfiguration is present.

As already mentioned, the approaches for building such a classifier as a detection approach, in general, can vary depending on the field of application. This is true mainly regarding the origin of the data. For the detection of a certain misconfigured control curve in a specific grid segment, the operational data collected at the transformer is used. The DeMaDs framework needs to be calibrated using the operational data of the respective substation for a period of up to 2 weeks. This calibration is done under the assumption of regular operation. During this time the collected substation data are used as samples of operation without malfunctions, whereas the corresponding erroneous samples are generated using grid simulation. These simulations are made possible through the use of the equivalent of smart meter data which is obtained through data mining. This data mining uses load estimation performed by an Artificial Neural Network (ANN).

Such an ANN is trained using dummy data created by power grid simulations. The synthetic data allow the ANN to infer the relationships between voltages and powers in the grid. Therefore, without full knowledge of the grid state, it can estimate the loads. After the calibration phase, each new sample collected is classified as either regular or erroneous. In the latter case, a malfunction is detected. In case the sample is deemed regular, the erroneous sample is generated in the same manner as before. This leaves the DeMaDs framework with data of a rolling window of about 2 weeks to accommodate possible changes in the grid. This transformer-level detection uses classic ML approaches due to the high resolution and high dimensionality of the substation data used. This approach is able to detect a misconfiguration in a certain grid segment. Followingly, it is grid specific. It can also point to the kind of misconfiguration occurring since for each use case a separate classifier can be built.

To be able to tell at which device exactly a misconfiguration is present, a detection on a device level is necessary. The framework can therefore also take the results of the decentralized detection approach into consideration.

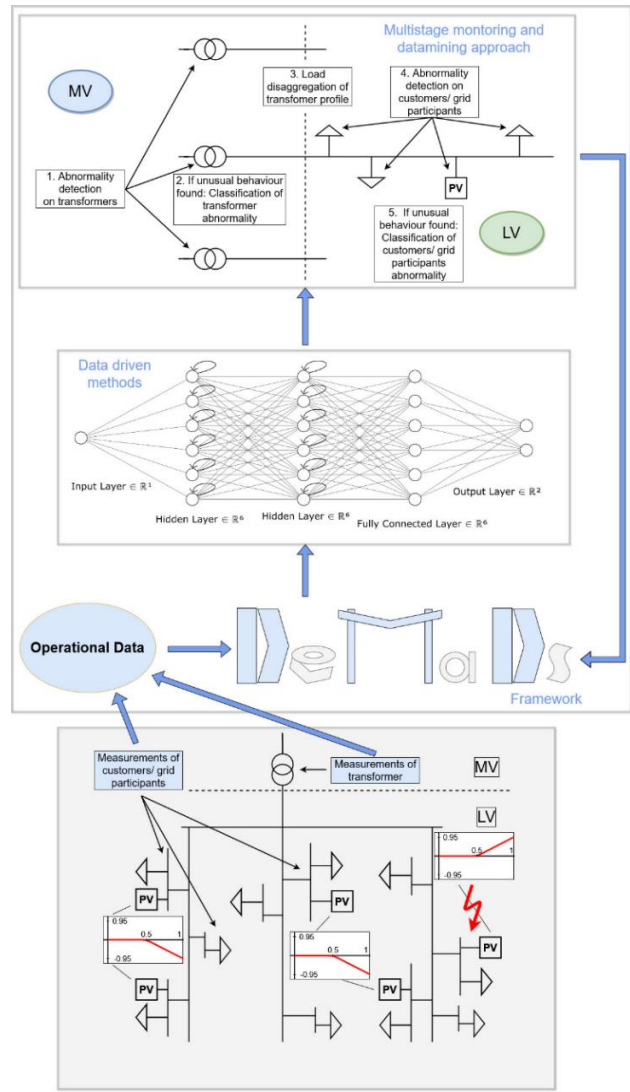

*Figure 1: Sketch of the functionality of the malfunction detection framework*

This approach, in contrast to the transformer-level approach, is grid unspecific. It works by pre-training a Deep Neural Network (DNN) by feeding it with meter data sourced from simulating various grids. This operational data also contains cases of grid operation while a malfunction is present, and cases of regular grid operation. The DL method applied here extracts the fundamental features of regular or malfunctioning grid operation from smart meter data. This allows the so-created model to recognize either case later on, independent of external factors such as the grid topology. Then, the pre-trained ANNs are rolled out decentrally at each metering point. Due to the local installation in this manner, the usage of meter data is possible to detect malfunctions. Again, a distinct ANN model is pre-trained for each use case of malfunction of a certain device. In case one of these models detects a specific malfunction, the framework can raise a flag at a certain location. This localizes the malfunction within the grid segment identified before. In addition, the local detection approach adds increased certainty about the accuracy of the transformer detection by either confirming or contradicting it. This applies both to the statement on the occurrence of a malfunction in general, but also to the classification of its type.
Moreover, the exact location of the malfunction can be determined while respecting data protection requirements by transmitting only the information about the result flag raised by the local detection approach.
The approach of combining a grid-specific transformer-level approach based on traditional ML with a local, universally applicable DL solution allows DSOs to implement the envisioned monitoring. The framework enables this monitoring to be implemented without tedious customizations or specific training for a DSO's control room staff. The advantage is rooted in how the classifiers are created: for the transformer-level approach, only a calibration period of a certain length is necessary. During this time, data is collected autonomously. In addition, the simulations and load estimation that are needed to form the complete dataset used to build the classifier are automatically conducted as well. After this period, the detection is operational and does not require further intervention. Similarly effortless for the system operator is the set-up and operation of the local detection approach: due to the pre-training of the ANNs, solely they need to be rolled out. These pre-trained models also only need to make predictions using the newly collected data, keeping the requirements on computing power low at their point of application. Nevertheless, the incorporation of new use cases, meaning the detection of new misconfigurations requires an update of the framework. However, this does not require anything else than the modeling of this malfunction and a general roll-out of the updated version of the framework. This makes the solution presented easily scalable. The framework does not need any further knowledge about grid codes or intended configurations. It works solely on operational data, detecting deviations from a state assumed to be desirable. The only requirement, therefore, is the availability of this operational data. This should be guaranteed by the sensors at substations and smart meters, which are already in place today. Therefore, neither upgrades on metering capabilities, nor major changes to the communication infrastructure should be necessary for a practical application of the framework. The application envisioned is the extension of DSOs Supervisory Control and Data Acquisition (SCADA) systems.

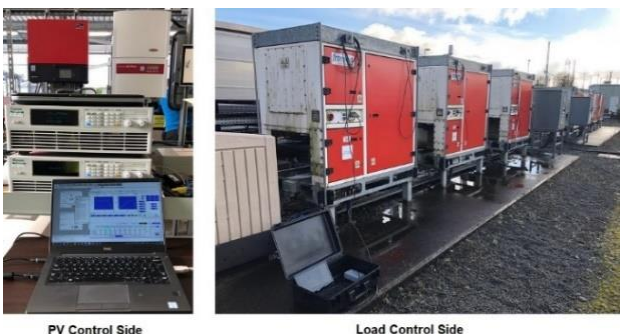

*Figure 2: Lab setups used for data collection*

## VALIDATION RESULTS

The outcomes of the evaluations on data properties and the data features used for detection are quite promising [6]. Moreover, a solution developed on data collected in a laboratory setting during an H2020 ERIGrid 2.0 Lab Access [7] showed very good results. The solution tested detected misconfigurations in almost all cases. These results and the results achieved on simulation data recreating the lab data are shown in Table 1. Figure 2 shows parts of the lab and the setup of load banks and the PV controls.

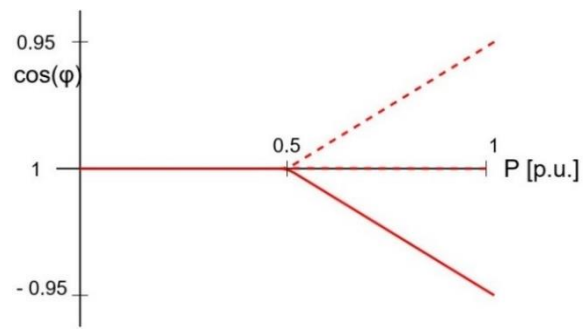

*Figure 3: Inverter configurations to be detected*

The validation for both the transformer detection approach as well as the detection approach on the device level was done on a PV inverter use case. The grid-supporting functionality to be monitored is a power factor control curve which, depending on the active power regulates the reactive power infeed of the inverter. This functionality helps to support the local voltage control. The control curve and its malfunctions modeled here are depicted in Figure 3: the fully drawn line is the correct control curve, whereas there are also two misconfiguration examples depicted. These are marked by dotted lines. The flat curve is called 'wrong' in the following, the inversed curve is simply marked as 'inversed'. The wrong control curve does not provide reactive power at all. The inverted curve feeds in reactive power of the opposite sign of the correct curve, which has the opposite effect of the intended voltage control.
The assessment of detection approaches on the device level revealed the so-called R Transformer approach to be the best-performing method [8]. This approach, which is sketched in Figure 4, uses Recurrent Neural Networks (RNN) to capture time dependencies in the data on a local scale. It also employs attention mechanisms to capture such dependencies globally over the entire time series [9]: Attention methods use so-called keys to encode the source data feature, which could be the features of a part of the entire input time series. Furthermore, queries are defined, which for example contain the hidden states connected to the last output. Through a score function, the relationship between queries and keys in deciding the next output is defined. This relationship is represented by the so-calculated energy score. This energy score reflects the importance of queries on the output, and thereby also the importance of the matched inputs encoded in the key.

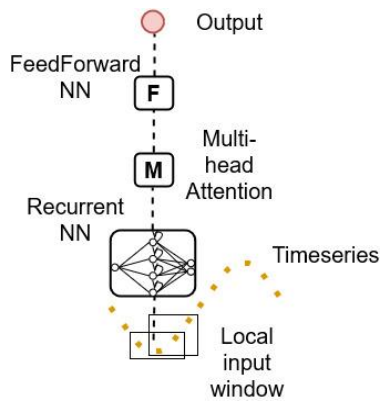

*Figure 4: R Transformer architecture*

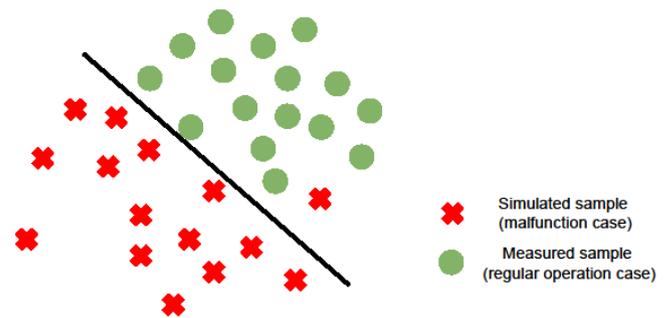

*Figure 5: Sketch of SVM classifier*

The architecture managed to detect misconfigurations grid unspecifically using simulated meter data. Other DL or traditional ML approaches did not satisfy this task. Therefore, it met the requirements set and was chosen as the best solution for the device-level detection part of the DeMaDs framework.

For the transformer-level detection approach, data were collected in a lab setting. This data was also reproduced using simulation. The experiments with this data identified the Support Vector Machine (SVM) to be the best-suited classifier here. Figure 5 illustrates the working principle of the SVM; the classifier is built by finding a decision boundary that separates the classes.

This decision boundary maximizes its margins meaning it keeps as big as possible distance to the samples.

The classes, as reflected by the results in Table 1 can encompass the correct operational state as well as one or more misconfigurations. The overall functionality of the complete monitoring application also becomes obvious; as already mentioned, the data of the last 14 days is measured. The data for each of these days is assumed to be correct. A day's measurements are condensed by data processing and are regarded as one sample.

Their malfunctioning counterparts are then simulated. At least one misconfiguration has to be modeled and simulated to have one class of a malfunctioning operational state. However, also both the wrong and inversed control curves mentioned before can be simulated, and therefore both serve as malfunction classes. The simulations are conducted after the substation measurements are disaggregated into its load components using load estimation. The expected PV production is assumed to be known by combining solar radiation models with the grid operators' knowledge about installed PV capacity. A newly measured sample is then classified as either measured under correct operation or while a malfunction was present. The data mining for the simulations, the simulations themselves, the dataset creation, and the entailing update on the classifier can be done centrally and only have to be conducted once a day. This keeps the computational cost low while still offering regular and reliable detection of malfunctions.

The assessments and validations were done with data reflecting properties that can be found in the measurements already available in the distribution grid. For the device-level detection approach, data in 15 minutes resolution was used. Furthermore, only voltage data was sufficient to detect the misconfigurations under scrutiny.

However, an extension to using also current measurements could become necessary to include other use cases. Current data, for example, is generally available at smart meters, which makes such a step unproblematic.

The transformer data detection was developed with data collected at a 4 Hertz frequency. This data is of high dimensionality, as it covers many variables such as voltages, currents as well as active and reactive power flows. The laboratory measurements were done phase-wise. This can yield advantages in a real-life application, depending on whether the devices to be monitored are installed in a single or three-phase manner.

The results of the transformer detection listed show that the approach has great potential. Both malfunctions, the misconfigured inversed and wrong control curves, were tried out in two different laboratory grid topologies. The measurements were re-created by simulation. In all cases, the detection performed very well, classifying the data almost always correctly meaning the detection was successful. As a result metric, the F-score is listed here; it balances Precision and Recall. Precision is defined by how many of the classifications done are actually accurately performed.

*Table 1: Detection results on PV misconfigurations [10]*

| F-Score Result Case | Grid Setup and Data Source | | | |
|---|---|---|---|---|
| | Grid Setup A | | Grid Setup B | |
| | Lab Data | Sim data | Lab Data | Sim data |
| correct vs. wrong | 0.93 | 0.91 | 1 | 0.90 |
| correct vs. inversed | * | 0.97 | 1 | 0.97 |
| correct vs. wrong vs. inversed | * | 0.88 | 0.96 | 0.90 |
| correct vs. abnormal | * | 0.95 | 1 | 0.95 |

*\*Not available due to lab access time limitations*

In contrast to this, Recall is a metric for how many of the malfunctions present were found. The F-score, therefore, reflects the practical usability for a grid operator of a method quite well. On one hand, a DSO would obviously want to find as many malfunctions as possible. On the other hand, the grid operator is very inclined not to respond to false alarms, since they can entail interventions by maintenance staff which are costly and waste resources. The results also show the performance of the detection if all malfunction types are summed under an 'abnormal' class. This task is also performed well by the detection with a score that appears to be averaging over the scores of the detection of the individual malfunctions. This implies, that the solution is very well applicable in combination with device-level detection. This detection would then offer further reliable information on the type of malfunction present.

## CONCLUSIONS

The new requirements regarding energy sustainability lead to new forms of generation and new consumers. These new grid participants or grid-connected devices are often installed decentrally in the power distribution system. Due to the historic development of the power system, this part of the power grid is designed to merely distribute electrical energy in a very static manner. Due to this decentralization of generation and the introduction of new devices and linked services in general, the operation of the electric power grid is becoming increasingly complex. To counter these new challenges, grid-connected devices have to provide grid-supporting functionalities. However, for the same legacy reasons, the distribution grid is ill-equipped with sensors. This makes the monitoring of the execution of the supported functionalities difficult. To ensure the grid is operated in a safe and reliable manner, DSOs need new solutions to tackle this problem.

The framework presented in the DeMaDs approach offers both capabilities testing and developing such solutions, but also validated monitoring approaches for certain use cases. The use cases include several misconfigurations of control curves of different devices, such as PV inverters. The benefits of the proposed approach are the remote and automatic detection of these misconfigurations with the sole use of operational data. Furthermore, no additional metering infrastructure is needed: substation metering data is used centrally, whereas meter data is used locally to avoid data protection issues. This allows for deployment and operation without specifically trained staff, making DeMaDs easily integrable into DSOs' existing SCADA systems.

Future work includes the modeling of additional use cases, to allow for the detection of additional malfunctions. Furthermore, new developments in DL architectures should be periodically revised for integration into the framework's local detection to keep up with recent developments in this fast-moving field. Lastly, a field test is envisioned to refine the approach and test its applicability and robustness on a broader scale.

## ACKNOWLEDGMENT

This work received funding from the Austrian Research Promotion Agency (FFG) under the "Research Partnerships – Industrial PhD Program" in DeMaDs (FFG No. 879017) and from the European Community's Horizon 2020 Program (H2020/2014-2020) in project "ERIGrid 2.0" (Grant Agreement No. 870620) under the Lab Access user project #115 at the Power Network Demonstration Center (PNDC) of the University of Strathclyde.

## REFERENCES

[1] O. Wagner, M. Venjakob, and J. Schröder, "The growing impact of decentralised actors in power generation: a comparative analysis of the energy transition in Germany and Japan," Journal of Sustainable Development of Energy, Water and Environment Systems, vol. 9, no. 4, pp. 1-22, 2021.

[2] A.S. Awad, D. Turcotte, and T. H. El-Fouly, "Impact assessment and mitigation techniques for high penetration levels of renewable energy sources in distribution networks: voltage-control perspective," Journal of Modern Power Systems and Clean Energy, vol. 10, no. 2, pp. 450-458, 2021.

[3] S. Wang, P. Dehghanian, L. Li, and B. Wang, "A Machine Learning Approach to Detection of Geomagnetically Induced Currents in Power Grids," IEEE Transactions on Industry Applications, vol. 56, no. 2, pp. 1098-1106, 2020.

[4] M. Yasinzadeh, M. Akhbari, "Detection of PMU spoofing in power grid based on phasor measurement analysis," IET Generation, Transmition & Distribution, vol. 12, no. 9, pp. 1980-1987, 2018.

[5] S. Barja-Martinez, et al., "Artificial intelligence techniques for enabling big data services in distribution networks: A review," Renewable and Sustainable Energy Reviews, vol. 150, p. 111459, 2021.

[6] D. Fellner, H. Brunner, T.I. Strasser, and W. Kastner, "Towards data-driven malfunctioning detection in public and industrial power grids," 2020 25th IEEE International Conference on Emerging Technologies and Factory Automation (ETFA), Vienna, Austria, pp. 1451-1454, 2020.

[7] ERIGrid 2.0 Lab Access programme, [Online] https://erigrid2.eu/, accessed 20.07.2022

[8] D. Fellner, T. I. Strasser, and W. Kastner, "Applying deep learning-based concepts for the detection of device misconfigurations in power systems," Sustainable Energy, Grids and Networks, vol. 32, p. 100851, 2022.

[9] Z. Niu, G. Zhong, and H. Yu, "A review on the attention mechanism of deep learning," Neurocomputing, vol. 452, pp. 48–62, 2021

[10] D. Fellner, T.I. Strasser, W. Kastner, B. Feizifar, and I.F. Abdulhadi, "Data Driven Transformer Level Misconfiguration Detection in Power Distribution Grids," 2022 IEEE SMC, Prague, Czech Republic, pp. 1840-1847, 2022.